\documentclass[12pt,english,draftcls, one column]{IEEEtran}
\usepackage[T1]{fontenc}
\usepackage[latin9]{inputenc}
\usepackage{float}
\usepackage{amsmath}
\usepackage{amsthm}
\usepackage{amssymb}
\usepackage{graphicx}
\usepackage{stfloats}
\usepackage{cite} 
\usepackage{xcolor}

\makeatletter

\floatstyle{ruled}
\newfloat{algorithm}{tbp}{loa}
\providecommand{\algorithmname}{Algorithm}
\floatname{algorithm}{\protect\algorithmname}

\theoremstyle{plain}

\theoremstyle{plain}

\ifCLASSINFOpdf% \usepackage[pdftex]{graphicx}
\else% or other class option (dvipsone, dvipdf, if not using dvips). graphicx
\fi% graphicx was written by David Carlisle and Sebastian Rahtz. It is
\usepackage{babel}

\makeatother

\usepackage{babel}
\providecommand{\propositionname}{Proposition}
\providecommand{\theoremname}{Theorem}

\begin{document}

\title{Robust Design of Rate-Splitting Multiple Access With Imperfect CSI for \\ Cell-Free MIMO Systems}

\author{Daesung Yu, \textit{Graduate Student Member}, \textit{IEEE}, Seok-Hwan Park, \textit{Member}, \textit{IEEE}, Osvaldo Simeone, \textit{Fellow}, \textit{IEEE}, and Shlomo Shamai (Shitz), \textit{Life Fellow}, \textit{IEEE}
\thanks{
This work was supported by the National Research Foundation (NRF) of Korea Grants funded by the Ministry of Science and ICT under Grant NRF-2021R1C1C1006557; in part by the Ministry of Education under Grant 2021R1A6A3A13046157.
This work was also supported by the European Research Council (ERC) under the European Union's Horizon 2020 Research and Innovation Programme under Grant 694630 and 725731.

D. Yu and S.-H. Park are with the Division of Electronic Engineering, Jeonbuk
National University, Jeonju 54896, Korea (email: imcreative93@jbnu.ac.kr, seokhwan@jbnu.ac.kr).

O. Simeone is with King's Communication, Learning and Information Processing (kclip) Lab, the Centre for Telecommunications Research, Department of Engineering, King's College London, London WC2R 2LS, U.K (email: osvaldo.simeone@kcl.ac.uk).

S. Shamai is with the Department of Electrical and Computer Engineering, Technion, Haifa 3200003, Israel (email: sshlomo@ee.technion.ac.il).

\copyright \, 2022 IEEE. Personal use of this material is permitted. Permission from IEEE must be obtained for all other uses, in any current or future media, including reprinting/republishing this material for advertising or promotional purposes, creating new collective works, for resale or redistribution to servers or lists, or reuse of any copyrighted component of this work in other
works.

}}
\maketitle
\begin{abstract}
Rate-Splitting Multiple Access (RSMA) for multi-user downlink operates by splitting the message for each user equipment (UE) into a private message and a set of common messages, which are simultaneously transmitted by means of superposition coding. 
The RSMA scheme can enhance throughput and connectivity as compared to conventional multiple access techniques by optimizing the rate-splitting ratios along with the corresponding downlink beamforming vectors.
This work examines the impact of erroneous channel state information (CSI) on the performance of RSMA in cell-free multiple-input multiple-output (MIMO) systems.
An efficient robust optimization algorithm is proposed by using closed-form lower bound expressions on the expected data rates.
Extensive numerical results show the importance of robust design in the presence of CSI errors and how the performance gain of RSMA over conventional schemes is affected by CSI imperfection.\end{abstract}


\theoremstyle{theorem}
\newtheorem{theorem}{Theorem}
\theoremstyle{proposition}
\newtheorem{proposition}{Proposition}
\theoremstyle{lemma}
\newtheorem{lemma}{Lemma}
\theoremstyle{corollary}
\newtheorem{corollary}{Corollary}
\theoremstyle{definition}
\newtheorem{definition}{Definition}
\theoremstyle{remark}
\newtheorem{remark}{Remark}

\section{Introduction} \label{sec:intro}

Rate-splitting multiple access (RSMA) \cite{Mao:EJWCN18, Mao:arxiv22} is a powerful transmission technique that generalizes two conventional multiple access schemes, space-division multiple access (SDMA) \cite{Christensen:TWC08} and non-orthogonal multiple access (NOMA) \cite{Hanif:TSP16} for multi-user downlink systems.
It achieves enhanced throughput and connectivity via interference management. 
In RSMA, the message intended for each user equipment (UE) is split into a private message and a set of common messages. They are encoded to independent codewords and simultaneously transmitted on the downlink channel with superposition coding.
Each UE decodes a subset of common messages with a successive interference cancellation (SIC) decoding as in NOMA, and lastly decodes its private message.
Reference \cite{Mao:EJWCN18} proposed a general form of RSMA strategy for multi-user downlink.
Follow-up work has studied the potential advantages of RSMA from different perspectives such as robust transmission with partial channel state information (CSI) \cite{Joudeh:TCOM16, Ahmad:TCOM21}, joint sensing and communication \cite{Xu:JSTSP21}, cell-free multiple-input multiple-output (MIMO) or cloud radio access network (C-RAN) \cite{Ahmad:TCOM21, Yu:WCL19}, and simultaneous wireless information and power transfer (SWIPT) systems \cite{Acosta:SJ21}.

In this work, we investigate the impact of imperfect CSI on the performance of RSMA in cell-free MIMO systems, in which a central processor (CP) serves UEs through distributed access points (APs) connected via finite-capacity fronthaul links \cite{Park:TSP13}.
We assume that the CP designs the downlink beamforming and fronthaul quantization strategies based on global instantaneous CSI, which is affected by estimation errors \cite{Joudeh:TCOM16, Pan:TWC18, Choi:TWC20}.
Specifically, the CP has knowledge of global estimated CSI and of the distribution of estimation errors.
For this scenario, we discuss robust designs of SDMA, NOMA, and RSMA. To facilitate stochastic optimization, we use Jensen's inequality to obtain closed-form lower bound expressions of the achievable rates. 
We tackle the problems of maximizing the minimum-rate metric under SDMA, NOMA and RSMA via the majorization minimization (MM) approach \cite{Park:TSP13}.
We present extensive numerical results that show the importance of robust design in the presence of CSI errors, as well as the impact of erroneous CSI on the performance of various multiple access schemes.

\section{System Model\label{sec:System-Model}}

\begin{figure}
\centering\includegraphics[width=0.8\linewidth]{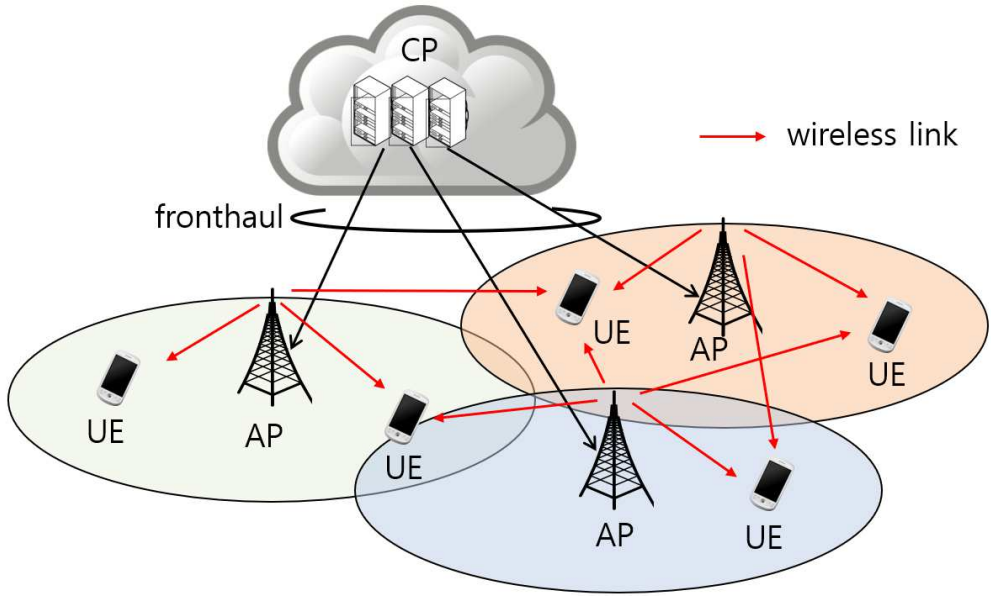}
\vspace{-3mm}
\caption{\label{fig:system-model}An example of cell-free MIMO system, in which a CP communicates with $K=5$ UEs through $M=3$ APs.}
\end{figure}

As shown in Fig. \ref{fig:system-model}, we consider the downlink of a cell-free MIMO system in which $K$ single-antenna UEs are served by $M$ single-antenna APs controlled by a CP. 
Define the index sets for APs and UEs as $\mathcal{M} = \{1,2,\ldots,M\}$ and $\mathcal{K} = \{1,2,\ldots,K\}$, respectively.
The discrete message intended for UE $k\in\mathcal{K}$ and its data rate in bps/Hz are denoted by $W_k$ and $R_k$, respectively.
The CP communicates with each AP $i\in\mathcal{M}$ via a digital fronthaul link of capacity $C_{\text{fh}}$ bps/Hz.

The received signal of UE $k\in\mathcal{K}$ is written as
\begin{align}
    y_k = \mathbf{h}_k^H \mathbf{x} + n_k, \label{eq:channel-model}
\end{align}
where $\mathbf{h}_k\in\mathbb{C}^{M\times 1}$ denotes the channel vector from all APs to UE $k$, $\mathbf{x}\in\mathbb{C}^{M\times 1}$ is the signal vector transmitted by all APs, and $n_k\sim\mathcal{CN}(0,\sigma^2)$ indicates the additive noise at UE $k$.
The $i$th elements of $\mathbf{h}_k$ and $\mathbf{x}$ are respectively denoted by $h_{k,i}$ and $x_i$.
Each AP $i$ has a limited power budget $P_{\text{tx}}$, imposing $\mathbb{E}[|x_i|^2]\leq P_{\text{tx}}$ for all $i\in\mathcal{M}$.
The channel coefficients $h_{k,i}$ are independent across the indices $k$ and $i$ due to the fact that the APs and UEs are distributed.
Also, we assume Rayleigh fading channels $h_{k,i} \sim \mathcal{CN}(0, \alpha_{k,i})$, where $\mathbb{E}[|h_{k,i}|^2] = \alpha_{k,i}$ stands for the path-loss between UE $k$ and AP $i$.

We consider a time-division duplexing (TDD) system, in which the downlink CSI is obtained by uplink channel training process.
Accordingly, the UEs send pilot signals on the uplink channel, and the APs transfer the received pilot signals to the CP which estimates global CSI based on the collected pilot signals. 
We denote an estimate of $h_{k,i}$ by $\hat{h}_{k,i}$, which is related to the true channel $h_{k,i}$ as
\begin{align}
    h_{k,i} = \hat{h}_{k,i} + e_{k,i}. \label{eq:additive-error-CSI-model}
\end{align}
Here the difference $e_{k,i} = h_{k,i} - \hat{h}_{k,i}$ indicates the estimation error.
If a linear minimum mean squared error (MMSE) estimator is employed at the CP, estimate $\hat{h}_{k,i}$ and error $e_{k,i}$ are uncorrelated, and $e_{k,i}$ follows a Gaussian distribution, i.e., $e_{k,i}\sim\mathcal{CN}(0, z_{k,i})$ \cite{Pan:TWC18, Choi:TWC20}, where $z_{k,i} = \mathbb{E}[|e_{k,i}|^2]$ represents the estimation error power.
Thus, we have $\alpha_{k,i} = \mathbb{E}[|\hat{h}_{k,i}|^2] + z_{k,i}$.

Throughout the paper, we assume that the CP has knowledge of global estimated CSI $\hat{\mathbf{h}} = \{\hat{h}_{k,i}\}_{k\in\mathcal{K}, i\in\mathcal{M}}$ and of the estimation error powers $\mathbf{z} = \{z_{k,i}\}_{k\in\mathcal{K}, i\in\mathcal{M}}$.
Also, for fairness among the UEs, we focus on maximizing the minimum-rate $R_{\min} = \min_{k\in\mathcal{K}} R_k$ of all UEs.

\section{Single-Layer Schemes: SDMA and NOMA}
\label{sec:single-layer}

In this section, we discuss single-layer transmission schemes, in which the message $W_k$ for each UE $k$ is encoded into a baseband data signal $s_k \sim \mathcal{CN}(0,1)$. 
Each signal $s_k$ is beamformed with a vector $\mathbf{v}_k\in\mathbb{C}^{M\times 1}$, and the beamforming output signals are quantized and compressed to bit streams which are delivered to the corresponding APs on fronthaul links. The transmitted signal vector $\mathbf{x}$ of APs, obtained by decompressing the fronthaul bit streams, is given as
\begin{align}
    \mathbf{x} = \sum_{k\in\mathcal{K}} \mathbf{v}_k s_k + \mathbf{q}. \label{eq:beamforming-single-layer}
\end{align}
In (\ref{eq:beamforming-single-layer}), the vector $\mathbf{q}\sim\mathcal{CN}(\mathbf{0}, \boldsymbol{\Omega})$ represents quantization distortion noise, which is uncorrelated with the beamformed signal $\tilde{\mathbf{x}} = \sum_{k\in\mathcal{K}} \mathbf{v}_k s_k$. 
We consider an independent quantization of the elements of $\tilde{\mathbf{x}} = [\tilde{x}_1 \, \tilde{x}_2 \, \cdots \, \tilde{x}_M]^T$ so that we have $\boldsymbol{\Omega} = \text{diag}(\{\omega_i\}_{i\in\mathcal{M}})$ \cite{Simeone:EJASP09}.
The beamforming vectors $\mathbf{v}=\{\mathbf{v}_k\}_{k\in\mathcal{K}}$ and the quantization noise powers $\boldsymbol{\omega} = \{\omega_i\}_{i\in\mathcal{M}}$ must satisfy the constraints
\begin{subequations} \label{eq:constraints-power-fronthaul-single}
\begin{align}
&\sum_{k\in\mathcal{K}}|v_{k,i}|^2 + \omega_i \leq P_{\text{tx}}, \, i\in\mathcal{M}, \label{eq:constraints-power-single} \\
&\omega_i \geq \beta \sum_{k\in\mathcal{K}}|v_{k,i}|^2, \, i\in\mathcal{M}, \label{eq:constraints-fronthaul-single}
\end{align}
\end{subequations}
where $v_{k,i}$ denotes the $i$th element of $\mathbf{v}_k$, and $\beta$ is defined as $\beta = 1/ (2^{C_{\text{fh}}} - 1)$.
The constraint (\ref{eq:constraints-power-single}) rewrites the power constraint $\mathbb{E}[|x_i|^2]\leq P_{\text{tx}}$, and the constraint (\ref{eq:constraints-fronthaul-single}) is imposed for successful decompression at APs and comes from the fronthaul capacity constraints $I(\tilde{x}_i; x_i)\leq C_{\text{fh}}$ \cite{Park:TSP13, Simeone:EJASP09}.
%We note that we can fix the quantization noise powers as $\omega_i = \beta \sum_{k\in\mathcal{K}} |v_{k,i}|^2$ without loss of optimality.

Each UE $k$ decodes its intended signal $s_k$ using the received signal $y_k = \sum_{l\in\mathcal{K}} \mathbf{h}_k^H\mathbf{v}_l s_l + \mathbf{h}_k^H \mathbf{q} + z_k$. 
The single-layer schemes with single-user decoding (SUD) \cite{Christensen:TWC08} and SIC decoding \cite{Hanif:TSP16} are referred to as SDMA and NOMA, respectively \cite{Mao:EJWCN18}.
We discuss these schemes in the following subsections.

\subsection{SDMA (Single-User Decoding)} \label{sub:SDMA}

In the SDMA scheme \cite{Mao:EJWCN18}, UE $k$ decodes the data signal $s_k$ while treating the interference signals $\{s_l\}_{l\in\mathcal{K}\setminus\{k\}}$ as noise.
For given estimated CSI $\hat{\mathbf{h}}_k$, the expected data rate $R_k$ of UE $k$  is given as \cite{Pan:TWC18}
\begin{align}
    R_k = \mathbb{E}_{\mathbf{e}_k}\left[ \log_2\left(1 + 
    \gamma_k \right) \Big| \hat{\mathbf{h}}_k \right], \label{eq:achievable-rate-expectation}
\end{align}
with the error vector $\mathbf{e}_k = [e_{k,1}\,e_{k,2}\,\cdots\,e_{k,M}]^T$ and the signal-to-interference-plus-noise ratio (SINR)
\begin{align}
    \gamma_k = \frac{ \hat{\mathbf{h}}_{k}^H \mathbf{V}_k\hat{\mathbf{h}}_{k} }{ \sigma^2 + \mathbf{e}_k^H \mathbf{V}_{k} \mathbf{e}_k + \mathbf{h}_k^H\left(\boldsymbol{\Omega} + \mathbf{V}_{\mathcal{K}\setminus\{k\}} \right)\mathbf{h}_k }. \label{eq:SINR-for-fixed-error}
\end{align}
In (\ref{eq:SINR-for-fixed-error}), we have defined the notations $\mathbf{V}_k = \mathbf{v}_k\mathbf{v}_k^H$ and $\mathbf{V}_{\mathcal{S}} = \sum_{m\in\mathcal{S}} \mathbf{V}_m$. The second term $\mathbf{e}_k^H \mathbf{V}_{k} \mathbf{e}_k$ in the denominator of (\ref{eq:SINR-for-fixed-error}) indicates the self-interference signal power caused by erroneous CSI.

As in \cite{Pan:TWC18, Choi:TWC20}, we use Jensen's inequality to obtain a closed-form expression for (\ref{eq:SINR-for-fixed-error}), which serves as a lower bound on the rate (\ref{eq:achievable-rate-expectation}), i.e.,
\begin{align}
    R_k \geq f_{k}(\mathbf{V}, \boldsymbol{\omega}) = \log_2\left( 1 + \tilde{\gamma}_k \right), \label{eq:achievable-rate-lower-bound-SDMA}
\end{align}
where $\tilde{\gamma}_k$ is defined as
\begin{align}
    \tilde{\gamma}_k &= \frac{\hat{\mathbf{h}}_{k}^H \mathbf{V}_k\hat{\mathbf{h}}_{k}}{ \sigma^2 + \mathbb{E}_{\mathbf{e}_k}\big[\mathbf{e}_k^H \mathbf{V}_{k} \mathbf{e}_k\big] + \mathbb{E}_{\mathbf{e}_k}\big[\mathbf{h}_k^H\left(\boldsymbol{\Omega} + \mathbf{V}_{\mathcal{K}\setminus\{k\}} \right)\mathbf{h}_k \big| \hat{\mathbf{h}}_k \big] } \nonumber \\
    & = \frac{ \hat{\mathbf{h}}_{k}^H \mathbf{V}_k\hat{\mathbf{h}}_{k} }{ \sigma^2 + \mathbf{z}_k^H (\boldsymbol{\Omega} + \mathbf{V}_{d, \mathcal{K}} ) \mathbf{z}_k + \hat{\mathbf{h}}_k^H\left(\boldsymbol{\Omega} + \mathbf{V}_{\mathcal{K}\setminus\{k\}} \right)\hat{\mathbf{h}}_k },
\end{align}
with the notation $\mathbf{z}_k=[z_{k,1}^{1/2}\,z_{k,2}^{1/2}\,\cdots,z_{k,M}^{1/2}]^T$ and the matrix $\mathbf{V}_{d,\mathcal{S}}$ obtained by replacing all the off-diagonal elements of $\mathbf{V}_{\mathcal{S}}$ with zeros.

The optimization of beamforming $\mathbf{V}$ and fronthaul quantization noise powers $\boldsymbol{\omega}$ of SDMA for minimum-rate maximization can be formulated as
\begin{subequations} \label{eq:problem-single}
\begin{align}
    \underset{\mathbf{V}, \boldsymbol{\omega}} {\mathrm{max.}}\,\,\, & \min_{k\in\mathcal{K}} f_{k}(\mathbf{V}, \boldsymbol{\omega}) \label{eq:problem-single-cost} \\
 \mathrm{s.t. }\,\,\,\,\,\, & \mathbf{e}_i^H\mathbf{V}_{\mathcal{K}}\mathbf{e}_i + \omega_i \leq P_{\text{tx}}, \, i\in\mathcal{M} \label{eq:problem-single-power} \\
 & \omega_i \geq \beta \mathbf{e}_i^H\mathbf{V}_{\mathcal{K}}\mathbf{e}_i, \, i\in\mathcal{M} \label{eq:problem-single-fronthaul} \\
 & \text{rank}(\mathbf{V}_k) \leq 1, \, k\in\mathcal{K},\label{eq:problem-single-rank}
\end{align}
\end{subequations}
where $\mathbf{e}_i\in\mathbb{C}^{M\times 1}$ is a vector whose elements are zeros except for the $i$th element equal to 1.
The problem (\ref{eq:problem-single}) can be converted to a difference-of-convex (DC) problem by rewriting the function $f_k(\mathbf{\mathbf{V}}, \boldsymbol{\omega})$ in (\ref{eq:achievable-rate-lower-bound-SDMA}) as
\begin{align}
    &\log_2\left( \sigma^2 + \mathbf{z}_k^H\left(\boldsymbol{\Omega} + \mathbf{V}_{d,\mathcal{K}}\right)\mathbf{z}_k + \hat{\mathbf{h}}_k^H \left( \boldsymbol{\Omega} + \mathbf{V}_{\mathcal{K}} \right) \hat{\mathbf{h}}_k \right) \nonumber \\
    &- \log_2\left( \sigma^2 + \mathbf{z}_k^H\left(\boldsymbol{\Omega} + \mathbf{V}_{d,\mathcal{K}}\right)\mathbf{z}_k + \hat{\mathbf{h}}_k^H \left( \boldsymbol{\Omega} + \mathbf{V}_{\mathcal{K}\setminus\{k\}} \right) \hat{\mathbf{h}}_k \right), \nonumber
\end{align}
and relaxing the rank constraint (\ref{eq:problem-single-rank}).
Thus, we can find a locally optimal solution by adopting the MM approach \cite{Park:TSP13} of the modified problem, and a feasible rank-1 solution can be found by a projection process.

\subsection{NOMA (SIC Decoding)} \label{sub:NOMA}

In the standard downlink NOMA scheme \cite{Hanif:TSP16}, the UEs perform SIC decoding with a given decoding order $s_{\pi(1)}\rightarrow s_{\pi(2)} \rightarrow \ldots \rightarrow s_{\pi(K)}$, until their intended signals are obtained.
With this approach, each data signal $s_{\pi(k)}$ should be successfully decoded at the intended UE $\pi(k)$ as well as the UEs $\pi(k+1),\pi(k+2),\ldots,\pi(K)$ which decode it before their intended signals.
Thus, the expected data rate $R_{\pi(k)}$ of UE $\pi(k)$ under NOMA is given as
\begin{align}
    R_{\pi(k)} = \min_{l\in[k:K]} \mathbb{E}_{\mathbf{e}_{\pi(l)}} \left[ \log_2\left(1 + \gamma_{l,k}\right) \Big| \hat{\mathbf{h}}_{\pi(l)} \right], \label{eq:achievable-rate-NOMA}
\end{align}
where $[k:K] = \{k,k+1,\ldots,K\}$ and
\begin{align}
    & \gamma_{l,k} = \frac{\hat{\mathbf{h}}_{\pi(l)}^H \mathbf{V}_{\pi(k)} \hat{\mathbf{h}}_{\pi(l)}}{ \sigma^2 + \mathbf{e}_{\pi(l)}^H\mathbf{V}_{\pi(k)}\mathbf{e}_{\pi(l)} +  \mathbf{h}_{\pi(l)}^H\left( \boldsymbol{\Omega} + \mathbf{V}_{\{\pi(m)\}_{m=k+1}^K} \right) \mathbf{h}_{\pi(l)} }.
\end{align}

Similar to SDMA, we replace each expected rate $\mathbb{E}_{\mathbf{e}_{\pi(l)}} [ \log_2(1 + \gamma_{l,k}) \big| \hat{\mathbf{h}}_{\pi(l)} ]$ with the closed-form lower bound expression
\begin{align}
    f_{l,k}(\mathbf{V}, \boldsymbol{\omega}) = 
    \log_2\left( 1 + \frac{ \hat{\mathbf{h}}_{\pi(l)}^H \mathbf{V}_{\pi(k)} \hat{\mathbf{h}}_{\pi(l)}
     }{\sigma^2 + I_{N,l,k}}  \right), \label{eq:MI-NOMA}
\end{align}
where
\begin{align}
     I_{N,l,k} & = \mathbb{E}_{\mathbf{e}_{\pi(l)}}\left[ \mathbf{e}_{\pi(l)}^H \mathbf{V}_{\pi(k)} \mathbf{e}_{\pi(l)}\right] + \mathbb{E}_{\mathbf{e}_{\pi(l)}}\left[\mathbf{h}_{\pi(l)}^H\left( \boldsymbol{\Omega} + \mathbf{V}_{\{\pi(m)\}_{m=k+1}^K} \right) \mathbf{h}_{\pi(l)} \Big| \hat{\mathbf{h}}_{\pi(l)} \right]  \nonumber \\
     & = \mathbf{z}_{\pi(l)}^H\left(\boldsymbol{\Omega} + \mathbf{V}_{d,\mathcal{K}}\right)\mathbf{z}_{\pi(l)} + \hat{\mathbf{h}}_{\pi(l)}^H\left( \boldsymbol{\Omega} + \mathbf{V}_{\{\pi(m)\}_{m=k+1}^K} \right) \hat{\mathbf{h}}_{\pi(l)}.
\end{align}

We can state the problem of minimum-rate maximization for NOMA scheme as
\begin{subequations} \label{eq:problem-NOMA}
\begin{align}
    \underset{\mathbf{V}, \boldsymbol{\omega}, \mathbf{R}} {\mathrm{max.}}\,\,\, & \min_{k\in\mathcal{K}} R_k \label{eq:problem-NOMA-cost} \\
 \mathrm{s.t. }\,\,\,\,\,\, & R_{\pi(k)} \leq f_{l,k}(\mathbf{V}, \boldsymbol{\omega}), \, k\in\mathcal{K}, l\in[k:K] \label{eq:problem-NOMA-rate} \\
 & \mathbf{e}_i^H\mathbf{V}_{\mathcal{K}}\mathbf{e}_i + \omega_i \leq P_{\text{tx}}, \, i\in\mathcal{M} \label{eq:problem-NOMA-power} \\
 & \omega_i \geq \beta \mathbf{e}_i^H\mathbf{V}_{\mathcal{K}}\mathbf{e}_i, \, i\in\mathcal{M} \label{eq:problem-NOMA-fronthaul} \\
 &\text{rank}(\mathbf{V}_k)\leq 1, \, k\in\mathcal{K},\label{eq:problem-NOMA-rank}
\end{align}
\end{subequations}
with $\mathbf{R} = \{R_k\}_{k\in\mathcal{K}}$.
Similar to (\ref{eq:problem-single}), we can obtain a problem of DC form by rewriting the function $f_{l,k}(\mathbf{v}, \boldsymbol{\omega})$ in (\ref{eq:MI-NOMA}) as
\begin{align}
    &\log_2\left( \sigma^2 + I_{N,l,k} +  \hat{\mathbf{h}}_{\pi(l)}^H \mathbf{V}_{\pi(k)} \hat{\mathbf{h}}_{\pi(l)} \right) - \log_2\left( \sigma^2 + I_{N,l,k} \right), \nonumber
\end{align}
and removing the rank constraint (\ref{eq:problem-NOMA-rank}).

\section{Rate-Splitting Multiple Access} \label{sec:RSMA}

To describe a general RSMA scheme, we denote the number of common signals by $L$ and define the set of common signals' indices as $\mathcal{L} = \{1,2,\ldots,L\}$. 
We assume that the $l$th common signal is decoded by the UEs in subset $\mathcal{S}_l \subseteq \mathcal{K}$.
To remove redundancy in the design, the subsets $\mathcal{S}_1,\mathcal{S}_2,\ldots,\mathcal{S}_L$ satisfy the conditions $\mathcal{S}_{l_1}\neq \mathcal{S}_{l_2}$ for all $l_1\neq l_2 \in\mathcal{L}$. Also, to differentiate from private signals each of which is decoded by a single UE, we have $|\mathcal{S}_l| \geq 2$, $l\in\mathcal{L}$.
For given  $\mathcal{S}_1,\mathcal{S}_2,\ldots,\mathcal{S}_L$, we can figure out the set $\mathcal{L}_k$ of common signals' indices that contain UE $k$, i.e., $\mathcal{L}_k = \{l\in\mathcal{L} | k\in\mathcal{S}_l\}$.

\subsection{Overall Operation and Beamforming Optimization} \label{sub:operation-RSMA}

In the RSMA scheme, the message $W_k$ for UE $k$ is split into $L_k+1$ messages, i.e., a private message $W_{p,k}$ and $L_k$ common messages $\{W_{c,k,l}\}_{l\in\mathcal{L}_k}$, where $L_k = |\mathcal{L}_k|$.
The private message $W_{p,k}$ is decoded only by UE $k$, and the common message $W_{c,k,l}$ is decoded by all UEs in the set $\mathcal{L}_k$.
Denoting the rates of $W_{p,k}$ and $W_{c,k,l}$ by $R_{p,k}$ and $R_{c,k,l}$, respectively, we can write $R_k = R_{p,k} + \sum_{l\in\mathcal{L}_k} R_{c,k,l}$.

The private messages $W_{p,k}$, $k\in\mathcal{K}$, are separately encoded, so the resulting private symbols $s_{p,k}\sim\mathcal{CN}(0,1)$ with $k\in\mathcal{K}$ are independent.
The common messages $W_{c,k,l}$, $k\in\mathcal{S}_l$, associated with the $l$th subset $\mathcal{S}_l$ are concatenated and encoded to a single common signal $s_{c,l}\sim\mathcal{CN}(0,1)$. 
Accordingly, the common signal $s_{c,l}$ is decoded by all UEs in $\mathcal{S}_l$, and each UE $k\in\mathcal{S}_l$ can extract its common message $W_{c,k,l}$ from $s_{c,l}$.

Denoting the beamforming vectors for signals $s_{p,k}$ and $s_{c,l}$ by $\mathbf{v}_{p,k}\in\mathbb{C}^{M\times 1}$ and $\mathbf{v}_{c,l}\in\mathbb{C}^{M\times 1}$, respectively, the transmitted signal vector $\mathbf{x}$ of APs is given as
\begin{align}
    \mathbf{x} = \sum_{k\in\mathcal{K}} \mathbf{v}_{p,k} s_{p,k} + \sum_{l\in\mathcal{L}} \mathbf{v}_{c,l} s_{c,l} + \mathbf{q}. \label{eq:beamforming-RSMA}
\end{align}
If we denote the $i$th elements of $\mathbf{v}_{p,k}$ and $\mathbf{v}_{c,l}$ by $v_{p,k,i}$ and $v_{c,l,i}$, the power and fronthaul capacity constraints at AP $i$ can be written as
\begin{subequations} \label{eq:constraints-power-fronthaul-RSMA}
\begin{align}
    &\sum_{k\in\mathcal{K}}|v_{p,k,i}|^2 + \sum_{l\in\mathcal{L}} |v_{c,l,i}|^2 + \omega_i \leq P_{\text{tx}}, \, i\in\mathcal{M}. \label{eq:power-constraint-RSMA-general} \\
    & \omega_i \geq \beta\left( \sum_{k\in\mathcal{K}}|v_{p,k,i}|^2 + \sum_{l\in\mathcal{L}} |v_{c,l,i}|^2 \right), \, i\in\mathcal{M}. \label{fronthaul-constraint-RSMA-general}
\end{align}
\end{subequations}

UE $k$ decodes the messages $W_{p,k}$ and $\{W_{c,k,l}\}_{l\in\mathcal{L}_k}$, which were split from the original message $W_k$, based on the received signal $y_k$.
This means that UE $k$ should decode the corresponding data signals $s_{p,k}$ and $\{s_{c,l}\}_{l\in\mathcal{L}_k}$. 
We assume that the SIC decoding is performed with the order $s_{c,\pi_k(1)}\rightarrow s_{c,\pi_k(2)} \rightarrow \ldots \rightarrow s_{c,\pi_k(L_k)}\rightarrow s_{p,k}$.
The decoding order $\pi_k$ is chosen guaranteeing that $|\mathcal{S}_{\pi_k(l_1)}| \geq |\mathcal{S}_{\pi_k(l_2)}|$ for all $l_1 < l_2 \in \{1,2,\ldots,L_k\}$ \cite{Mao:EJWCN18,Yu:WCL19}.
%We leave the optimization of $\pi_k$ for future work, and the discussion of this work can be applied for arbitrarily fixed $\pi_k$.

The expected data rates $\{R_{p,k}\}_{k\in\mathcal{K}}$ and $\{R_{c,k,l}\}_{k\in\mathcal{K}, l\in\mathcal{L}_k}$ are achievable if they satisfy the conditions
\begin{subequations} \label{eq:rate-constraint-RSMA}
\begin{align}
    &\sum\nolimits_{k\in\mathcal{S}_l} \!\! R_{c,k,l} \!\leq\!  \min_{k\in\mathcal{S}_l} \mathbb{E}_{\mathbf{e}_k} \!\!\left[ \log_2\!\left(1 + \gamma_{c,k,l}\right) \!\Big| \hat{\mathbf{h}}_{k} \right]\!, l\in\mathcal{L}, \label{eq:rate-constraint-common}\\
    &R_{p,k} \leq \mathbb{E}_{\mathbf{e}_k}\left[ \log_2\left(1 + \gamma_{p,k}\right) \Big| \hat{\mathbf{h}}_k \right], \,\, k\in\mathcal{K}, \label{eq:rate-constraint-private}
\end{align}
\end{subequations}
where
\begin{subequations} \label{eq:SINR-RSMA}
\begin{align}
    & \gamma_{c,k,l} = \frac{ \hat{\mathbf{h}}_{k}^H\mathbf{V}_{c,l}\hat{\mathbf{h}}_k }{ \sigma^2 + \mathbf{e}_k^H \mathbf{V}_{c,l} \mathbf{e}_k +  \mathbf{h}_k^H \big( \boldsymbol{\Omega} + \mathbf{V}_{p,\mathcal{K}} + \mathbf{V}_{c,\mathcal{L}\setminus\mathcal{L}_k} + \mathbf{V}_{c, \mathcal{Q}_{k,l} } \big) \mathbf{h}_k } \label{eq:SINR-RSMA-common} \\
    & \gamma_{p,k} = \frac{\hat{\mathbf{h}}_k^H \mathbf{V}_{p,k} \hat{\mathbf{h}}_k}{ \mathbf{e}_k^H \mathbf{V}_{p,k} \mathbf{e}_k + \mathbf{h}_k^H \big(  \boldsymbol{\Omega} + \mathbf{V}_{p,\mathcal{K}\setminus\{k\}} + \mathbf{V}_{c,\mathcal{L}\setminus\mathcal{L}_k} \big) \mathbf{h}_k  }, \label{eq:SINR-RSMA-private}
\end{align}
\end{subequations}
with $\mathcal{Q}_{k,l} = \{\pi_k(q)\}_{q= \pi_k^{-1}(l)+1 }^{L_k}$.

Similar to Sec. \ref{sec:single-layer}, we consider lower bound expressions $\mathbb{E}_{\mathbf{e}_k} [ \log_2 (1 + \gamma_{c,k,l} ) \big| \hat{\mathbf{h}}_{k} ] \geq f_{c,l}(\mathbf{V}, \boldsymbol{\omega})$ and $\mathbb{E}_{\mathbf{e}_k} [ \log_2 (1 + \gamma_{p,k} ) \big| \hat{\mathbf{h}}_k ] \geq f_{p,k}(\mathbf{V}, \boldsymbol{\omega})$, where we define the functions
\begin{subequations} \label{eq:MI-function-RSMA}
\begin{align}
    &f_{c,k,l}(\mathbf{V}, \boldsymbol{\omega}) = \log_2\left(1 + \frac{\hat{\mathbf{h}}_{k}^H\mathbf{V}_{c,l}\hat{\mathbf{h}}_k}{  \sigma^2 + I_{c,k,l}}\right), \label{eq:MI-function-common} \\
    &f_{p,k}(\mathbf{V}, \boldsymbol{\omega}) = \log_2\left( 1 + \frac{ \hat{\mathbf{h}}_k^H \mathbf{V}_{p,k} \hat{\mathbf{h}}_k }{ \sigma^2 + I_{p,k}  } \right). \label{eq:MI-function-private}
\end{align}
\end{subequations}
with
\begin{subequations} \label{eq:interference-power-RSMA}
\begin{align}
    I_{c,k,l} &= \mathbf{z}_k^H\left( \boldsymbol{\Omega} + \mathbf{V}_{d,c,\mathcal{L}} + \mathbf{V}_{d,p,\mathcal{K}} \right)\mathbf{z}_k +  \hat{\mathbf{h}}_k^H\left(\boldsymbol{\Omega} + \mathbf{V}_{p,\mathcal{K}} + \mathbf{V}_{c,\mathcal{L}\setminus\mathcal{L}_k} + \mathbf{V}_{c, \mathcal{Q}_{k,l} } \right)\hat{\mathbf{h}}_k, \label{eq:interference-power-RSMA-common} \\
    I_{p,k} &= \mathbf{z}_k^H\left( \boldsymbol{\Omega} + \mathbf{V}_{d,c,\mathcal{L}} + \mathbf{V}_{d,p,\mathcal{K}} \right)\mathbf{z}_k + \hat{\mathbf{h}}_k^H \left( \boldsymbol{\Omega} + \mathbf{V}_{p,\mathcal{K}\setminus\{k\}} + \mathbf{V}_{c,\mathcal{L}\setminus\mathcal{L}_k} \right) \hat{\mathbf{h}}_k. \label{eq:interference-power-RSMA-private}
\end{align}
\end{subequations}
Unlike Sec. \ref{sec:single-layer}, $\mathbf{V}$ denotes $\mathbf{V} = \{\mathbf{V}_{p,k}\}_{k\in\mathcal{K}} \cup \{\mathbf{V}_{c,l}\}_{l\in\mathcal{L}}$.

The minimum-rate maximization for the RSMA scheme is stated as
\begin{subequations} \label{eq:problem-RSMA}
\begin{align}
     \underset{\mathbf{V},\boldsymbol{\omega}, \mathbf{R}} {\mathrm{max.}}\,\,\, & \min_{k\in\mathcal{K}} \left(R_{p,k} + \sum_{l\in\mathcal{L}_k}R_{c,k,l}\right) \, \label{eq:problem-RSMA-cost} \\
 \mathrm{s.t. }\,\,\,\,\,\, & \sum_{k\in\mathcal{S}_l}R_{c,k,l} \leq f_{c,k^{\prime},l}(\mathbf{V},\boldsymbol{\omega}), \, l\in\mathcal{L},
 k^{\prime}\in\mathcal{S}_l \label{eq:problem-RSMA-rate-common} \\
 & R_{p,k} \leq f_{p,k}(\mathbf{V},\boldsymbol{\omega}), \, k\in\mathcal{K} \label{eq:problem-RSMA-rate-private} \\
 &\mathbf{e}_i^H\left( \mathbf{V}_{c,\mathcal{L}} + \mathbf{V}_{p,\mathcal{K}} \right)\mathbf{e}_i + \omega_i \leq P_{\text{tx}}, \, i\in\mathcal{M} \label{eq:problem-RSMA-power} \\
    & \omega_i \geq \beta \mathbf{e}_i^H\left( \mathbf{V}_{c,\mathcal{L}} + \mathbf{V}_{p,\mathcal{K}} \right)\mathbf{e}_i, \, i\in\mathcal{M} \label{problem-RSMA-fronthaul} \\
    & \text{rank}(\mathbf{V}_{c,l}) \!\leq\! 1, \text{rank}(\mathbf{V}_{p,k}) \!\leq\! 1, \, l\in\mathcal{L}, k\in\mathcal{K}, \label{eq:problem-RSMA-rank}
\end{align}
\end{subequations}
with the notations $\mathbf{R} = \{R_{p,k}\}_{k\in\mathcal{K}}\cup\{R_{c,k,l}\}_{k\in\mathcal{K}, l\in\mathcal{L}_k}$, $\mathbf{V}_{c,\mathcal{S}} = \sum_{l\in\mathcal{S}}\mathbf{V}_{c,l}$, and $\mathbf{V}_{p,\mathcal{S}} = \sum_{k\in\mathcal{S}}\mathbf{V}_{p,k}$.
We can convert the problem (\ref{eq:problem-RSMA}) into a DC form by rewriting the functions $f_{c,k^{\prime},l}(\mathbf{V},\boldsymbol{\omega})$ and $f_{p,k}(\mathbf{V},\boldsymbol{\omega})$ in (\ref{eq:problem-RSMA-rate-common}) and (\ref{eq:problem-RSMA-rate-private}) as $\log_2( \sigma^2 + I_{c,k^{\prime}, l} + \hat{\mathbf{h}}_{k^{\prime}}^H \mathbf{V}_{c,l} \hat{\mathbf{h}}_{k^{\prime}} ) - \log_2( \sigma^2 + I_{c,k^{\prime}, l} )$ and $\log_2( \sigma^2 + I_{p,k} + \hat{\mathbf{h}}_{k}^H \mathbf{V}_{p,k} \hat{\mathbf{h}}_{k} ) - \log_2( \sigma^2 + I_{p,k} )$, respectively.
Thus, a locally optimal solution of the problem obtained by removing the rank constraints in (\ref{eq:problem-RSMA-rank}) can be found by an MM-based iterative algorithm \cite{Park:TSP13}. Accordingly, at each iteration, the solution is updated to a better point by solving the convex problem obtained by convexifying the non-convex constraints (\ref{eq:problem-RSMA-rate-common}) and (\ref{eq:problem-RSMA-rate-private}).
The detailed algorithm procedure is described in Algorithm 1, where 
\begin{subequations} \label{eq:convexified-functions}
\begin{align}
    \tilde{f}_{c,k^{\prime},l}(\mathbf{V}, \boldsymbol{\omega}, \mathbf{V}^{[t-1]}, \boldsymbol{\omega}^{[t-1]}) &= \log_2\left(  \sigma^2  + I_{c,k^{\prime},l} + \hat{\mathbf{h}}_{k^{\prime}}^H \mathbf{V}_{c,l} \hat{\mathbf{h}}_{k^{\prime}} \right) \nonumber \\
    & - \log_2\left(\sigma^2 + I_{c,k^{\prime},l}^{[t-1]}\right) - 
    \frac{1}{\ln 2} \frac{ I_{c,k^{\prime},l} - I_{c,k^{\prime},l}^{[t-1]} }{\sigma^2 + I_{c,k^{\prime},l}^{[t-1]}},\label{eq:convexified-function-common} \\
    \tilde{f}_{p,k}(\mathbf{V}, \boldsymbol{\omega}, \mathbf{V}^{[t-1]}, \boldsymbol{\omega}^{[t-1]}) &= \log_2\left( \sigma^2 + I_{p,k} + \hat{\mathbf{h}}_{k}^H \mathbf{V}_{p,k} \hat{\mathbf{h}}_{k} \right) \nonumber \\
    & - \log_2\left( \sigma^2 + I_{p,k}^{[t-1]} \right) - \frac{1}{\ln 2}\frac{I_{p,k} - I_{p,k}^{[t-1]}}{\sigma^2 + I_{p,k}^{[t-1]}}. \label{eq:convexified-function-private}
\end{align}
\end{subequations}
After Algorithm 1 is completed, feasible beamforming vectors $\mathbf{v}_{c,l}$ and $\mathbf{v}_{p,k}$ can be obtained from the quadratic matrices $\mathbf{V}_{c,l}$ and $\mathbf{V}_{p,k}$ by a projection process.
For example, we obtain $\mathbf{v}_{c,l}$ as
$\mathbf{v}_{c,l}\leftarrow \mathbf{u}(\mathbf{V}_{c,l}) \lambda^{1/2}(\mathbf{V}_{c,l})$, where $\mathbf{u}_1(\cdot)$ and $\lambda_1(\cdot)$ take the principal eigenvector and eigenvalue of the input square matrix.

\begin{algorithm}
\caption{MM based algorithm for problem (\ref{eq:problem-RSMA})}

\textbf{\footnotesize{}1}\textbf{ Initialize}

\textbf{\footnotesize{}2}~~~Set $\mathbf{V}^{[1]}$ such that $\max_{i\in\mathcal{M}} \mathbf{e}_i^H\mathbf{V}_{\mathcal{K}}^{[1]}\mathbf{e}_i = \frac{1}{1+\beta}P_{\text{tx}}$;

\textbf{\footnotesize{}3}~~~Set $\omega_i^{[1]} \leftarrow \frac{\beta}{1 + \beta}\mathbf{e}_i^H\mathbf{V}_{\mathcal{K}}^{[1]}\mathbf{e}_i$, $i\in\mathcal{M}$;

\textbf{\footnotesize{}4}~~~Set $t\leftarrow 1$;

\textbf{\footnotesize{}5}\textbf{ repeat}

\textbf{\footnotesize{}6}~~~Update $t\leftarrow t+1$;

\textbf{\footnotesize{}7}~~~Update $(\mathbf{V}^{[t]},\boldsymbol{\omega}^{[t]})$
as a solution of the problem (\ref{eq:problem-RSMA}) with the functions $f_{c,k^{\prime},l}(\mathbf{V},\boldsymbol{\omega})$ and

~~~~$f_{p,k}(\mathbf{V},\boldsymbol{\omega})$ replaced with $\tilde{f}_{c,k^{\prime},l}(\mathbf{V}, \boldsymbol{\omega}, \mathbf{V}^{[t-1]}, \boldsymbol{\omega}^{[t-1]})$ and $\tilde{f}_{p,k}(\mathbf{V}, \boldsymbol{\omega}, \mathbf{V}^{[t-1]}, \boldsymbol{\omega}^{[t-1]})$;

\textbf{\footnotesize{}8} \textbf{until} $|R_{\min}^{[t]}-R_{\min}^{[t-1]}|\leq\epsilon$;
\end{algorithm}

\subsection{Design of Common Signal Sets} \label{sub:design-common-signal-subsets}

If we choose $L=0$, i.e., no common signal, the RSMA scheme in this section reduces to the SDMA scheme discussed in Sec. \ref{sub:SDMA}. 
Thus, we assume that $L\geq 1$ and need to determine the subsets $\mathcal{S}_1, \mathcal{S}_2, \ldots, \mathcal{S}_L$.
In principle, there can be at most $L=2^K-1-K$ possible subsets for common signals. If we optimize RSMA utilizing all $2^K-1-K$ common signals, the problem size exponentially increases with $K$ due to the beamforming variables $\mathbf{V}$ leading to prohibitive optimization complexity.
Therefore, we need to keep $L$ small and choose good subsets $\mathcal{S}_1, \mathcal{S}_2, \ldots, \mathcal{S}_L$ as a function of the CSI.
It was proposed in \cite{Mao:SPAWC18} to use only a single common signal, i.e., $L=1$, decoded by all the UEs, i.e., $\mathcal{S}_1 = \mathcal{K}$.
With this choice, the optimization of beamforming variables $\mathbf{V}$ is more tractable than the above case with $L=2^K-1-K$, since the problem size linearly increases with $K$.
It was shown in \cite{Mao:SPAWC18, Yu:WCL19} that this scheme can provide notable gains over conventional single-layer schemes.

Reference \cite{Yu:WCL19} reported that the performance of RSMA can be further improved by exploiting $L=K-1$ common signals with carefully chosen subsets $\mathcal{S}_1, \mathcal{S}_2, \ldots, \mathcal{S}_L$. 
Note that the problem size for optimizing the beamforming variables $\mathbf{V}$ for this scheme still linearly increases with $K$.
Since the data signal $s_{c,l}$ is multicast to the UEs in $\mathcal{S}_l$ with a common beamforming vector $\mathbf{v}_{c,l}$, it is desirable to cluster UEs such that the UEs in the same set $\mathcal{S}_l$ have channel vectors of similar directions to minimize the beamforming loss caused by the direction mismatch among UEs.
Based on this observation, the work \cite{Yu:WCL19} proposed a hierarchical clustering algorithm based on a pairwise dissimilarity metric $d_{k,m}$ defined as
\begin{align}
    d_{k,l} = 1 - \frac{|\mathbf{h}_k^H \mathbf{h}_l|}{||\mathbf{h}_k|| ||\mathbf{h}_l||}. \label{eq:dissimilarity-metric}
\end{align}
The metric $d_{k,l}\in[0,1]$ approaches 0 as the channel vectors $\mathbf{h}_k$ and $\mathbf{h}_l$ are better aligned, while it equals 1 when $\mathbf{h}_k \perp \mathbf{h}_l$.
Since we assume imperfect CSI in this work, we compute the dissimilarity metric $d_{k,l}$ in (\ref{eq:dissimilarity-metric}) using the estimated channel vectors $\hat{\mathbf{h}}_{k}$ and $\hat{\mathbf{h}}_l$.

The hierarchical clustering algorithm operates by building a dendrogram \cite{Rokach:Springer05} in the bottom-up direction. In the initial bottom layer, there are $K$ nodes each corresponding to a UE. Then, we perform $K-1$ merging steps based on the complete-linkage distance $d_{\mathcal{S}_A,\mathcal{S}_B}$ between two clusters $\mathcal{S}_A$ and $\mathcal{S}_B$ defined as
\begin{align}
    d_{\mathcal{S}_A,\mathcal{S}_B} = \max_{k\in\mathcal{S}_A,m\in\mathcal{S}_B} d_{k,m}. \label{eq:complete-linkage-distance}
\end{align}
In the clustering algorithm, we exclude single-cardinality subsets and include whole-UE set in order to distinguish from private signals and guarantee at least the performance of the RSMA scheme in \cite{Mao:SPAWC18}.
It was shown in \cite{Yu:WCL19} that with the described clustering algorithm, we always obtain $L=K-1$ subsets. The detailed clustering algorithm is summarized in Algorithm 2.
\begin{algorithm}
\caption{Hierarchical clustering algorithm for determining subsets $\mathcal{S}_1,\mathcal{S}_2, \ldots, \mathcal{S}_{K-1}$}

\textbf{\footnotesize{}1}\textbf{ Initialize:} $\underbrace{\mathcal{C}_k\leftarrow \{k\}, k\in\mathcal{K}}_{\text{Bottom clusters}}$, $\underbrace{ \mathcal{J}\leftarrow\mathcal{K}   }_{\text{Clusters' indices}}$, $\underbrace{l\leftarrow 1}_{\text{Stage index}}$;

\textbf{\footnotesize{}2}\textbf{ repeat}

\textbf{\footnotesize{}3}~~~~Find $(k_1^*,k_2^*)\leftarrow \arg\min_{k_1,k_2\in\mathcal{J}, k_1\neq k_2} d_{\mathcal{C}_{k_1},\mathcal{C}_{k_2}}$;

\textbf{\footnotesize{}4}~~~~Merge $\mathcal{C}_{k_1^*} \leftarrow \mathcal{C}_{k_1^*} \cup \mathcal{C}_{k_2^*}$ updating $\mathcal{J}\leftarrow \mathcal{J}\setminus\{k_2^*\}$;

\textbf{\footnotesize{}5}~~~~Set $\mathcal{S}_l \leftarrow \mathcal{C}_{k_1^*}$ and $l\leftarrow l+1$;

\textbf{\footnotesize{}6} \textbf{until} $|\mathcal{J}| = 1$;
\end{algorithm}

\section{Numerical Results} \label{sec:numerical}

In this section, we investigate the performance of RSMA schemes for cell-free MIMO system with imperfect CSI via numerical results. 
We assume that the positions of APs and UEs are sampled from independent uniform distribution within a circular region of radius 100 m. 
Denoting the distance between AP $i$ and UE $k$ by $D_{k,i}$, we model the path-loss $\alpha_{k,i} = \mathbb{E}[|h_{k,i}|^2] = (D_{k,i}/D_{\text{ref}})^{-\eta}$ with the reference distance $D_{\text{ref}} = 30$ m and path-loss exponent $\eta=3$.
We define the relative CSI error $\rho_z \in [0,1]$ such that the CSI error variance $\mathbb{E}[|e_{k,i}|^2] = z_{k,i}$ equals $z_{k,i} = \rho_z \alpha_{k,i}$ for all $k\in\mathcal{K}$ and $i\in\mathcal{M}$. Accordingly, $\mathbb{E}[|\hat{h}_{k,i}|^2] = (1-\rho_z) \alpha_{k,i}$.
It is noted that the perfect CSI case can be captured with $\rho_z = 0$, while $\rho_z=1$ indicates that no CSI is available.
For all simulations, we consider overloaded cellular scenarios with $K > M$, in which a limited number of APs serve a larger number of Internet-of-Things (IoT) devices.

\begin{figure}
\centering\includegraphics[width=0.7\linewidth]{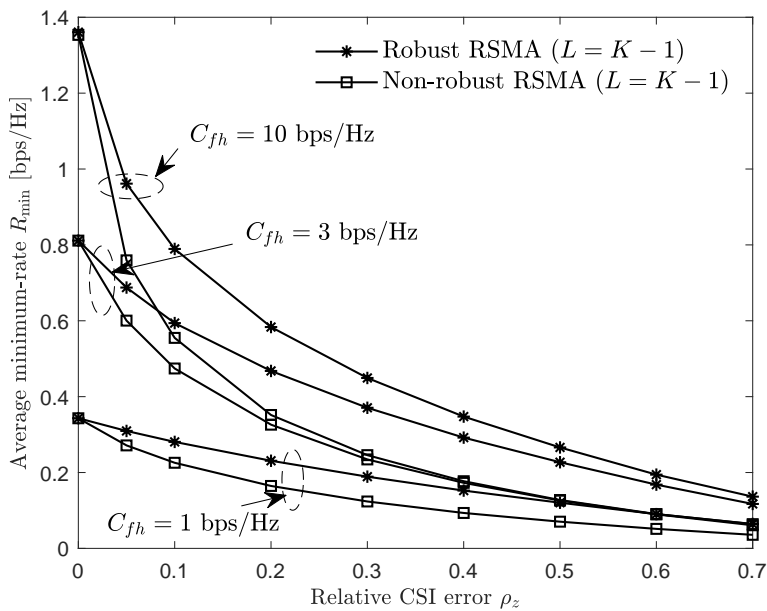}
%\vspace{-3mm}
\caption{\label{fig:vsRhoZ-robustness}Average minimum-rate $R_{\min}$ versus the relative CSI error $\rho_z$ ($M=4$, $K=8$, $C_{\text{fh}} \in \{1, 3, 10\}$ bps/Hz, and  $P_{\text{tx}}/\sigma^2 = 20$ dB)}
\end{figure}

In Fig. \ref{fig:vsRhoZ-robustness}, we examine the importance of robust design by plotting the performance of robust and non-robust RSMA schemes with $L=K-1$ common signals, while increasing the relative CSI error $\rho_z$ for a cell-free MIMO system with $M=4$, $K=8$, $C_{\text{fh}} \in \{1, 3, 10\}$ bps/Hz, and $P_{\text{tx}}/\sigma^2 = 20$ dB.
The non-robust scheme operates with the beamforming vectors $\mathbf{v}$ and fronthaul quantization noise powers $\boldsymbol{\omega}$ designed assuming no CSI error, i.e., $\rho_z=0$.
For both robust and non-robust schemes, we use $L=K-1$ common signals whose subsets $\mathcal{S}_1,\mathcal{S}_2,\ldots,\mathcal{S}_L$ are chosen with the hierarchical clustering method described in Algorithm 2.
The two schemes show the same performance when $\rho_z=0$, and the gap increases with $\rho_z$ in the regime of small $\rho_z$.
However, beyond a threshold level, the gap decreases with $\rho_z$, since the two schemes will show the same performance when $\rho_z=1$, which corresponds to the case of no CSI.
Comparing the cases of $C_{\text{fh}}\in\{1,3,10\}$ bps/Hz, the importance of robust design is seen to be more pronounced for a larger $C_{\text{fh}}$. 
This is because for small $C_{\text{fh}}$, the performance is more affected by the fronthaul quantization noise.

\begin{figure}
\centering\includegraphics[width=0.7\linewidth]{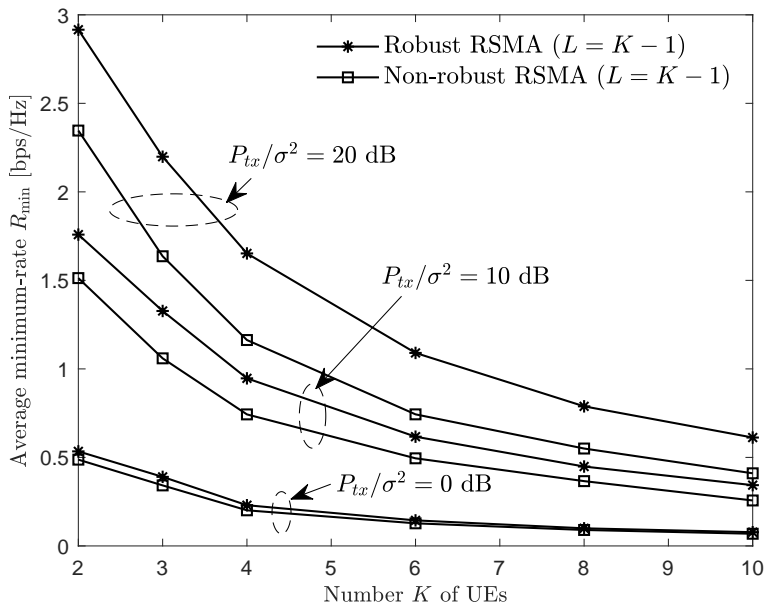}
%\vspace{-3mm}
\caption{\label{fig:vsK-robustness}Average minimum-rate $R_{\min}$ versus the number $K$ of UEs ($M=4$, $C_{\text{fh}} = 10$ bps/Hz, $P_{\text{tx}}/\sigma^2 \in \{0, 10, 20\}$ dB, and $\rho_z = 0.1$)}
\end{figure}

Fig. \ref{fig:vsK-robustness} plots the average minimum-rate $R_{\min}$ of the same schemes with respect to the number $K$ of UEs for a cell-free MIMO system with $M=4$, $C_{\text{fh}} = 10$ bps/Hz, $P_{\text{tx}}/\sigma^2 \in \{0, 10, 20\}$ dB, and $\rho_z = 0.1$. The performance of both schemes is degraded as $K$ increases, since the minimum-rate $R_{\min}$ is limited by the channel state of the worst UE.
For this reason, the performance gain of robust design becomes negligible when sufficiently many UEs are involved.
It is noted that as the transmit signal-to-noise ratio (SNR) $P_{\text{tx}}/\sigma^2$ increases, the gap between the two schemes becomes more significant.
This means that the impact of enhanced interference management with robust design is more evident when the additive noise power is smaller.

\begin{figure}
\centering\includegraphics[width=0.7\linewidth]{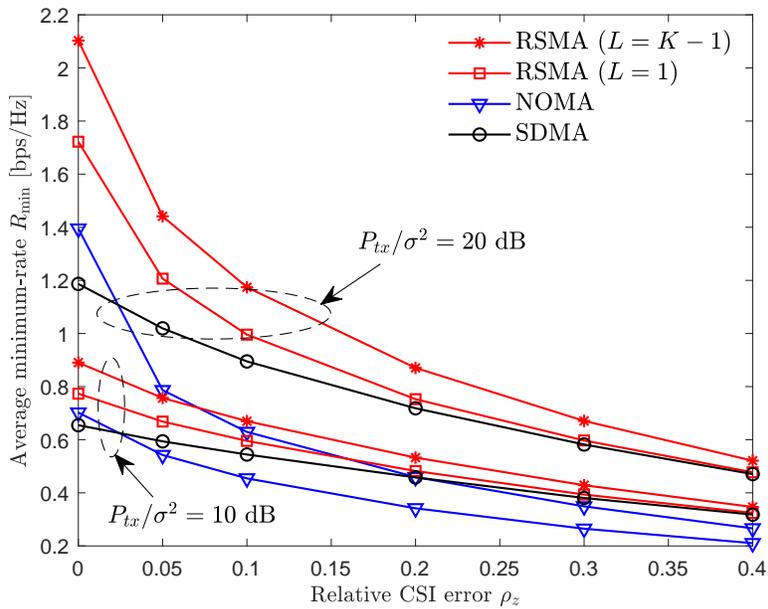}
%\vspace{-3mm}
\caption{\label{fig:vsRhoZ-various}Average minimum-rate $R_{\min}$ versus the relative CSI error $\rho_z$ ($M=4$, $K=6$, $C_{\text{fh}} = 10$ bps/Hz, and $P_{\text{tx}}/\sigma^2 \in \{10, 20\}$ dB)}
\end{figure}

In Fig. \ref{fig:vsRhoZ-various}, we compare the performance of various multiple access schemes: SDMA (Sec. \ref{sub:SDMA}), NOMA (Sec. \ref{sub:NOMA}), RSMA with $L=1$ common signal ($\mathcal{S}_1 = \mathcal{K}$) \cite{Mao:SPAWC18}, and RSMA with $L=K-1$ common signals ($\mathcal{S}_1,\mathcal{S}_2,\ldots,\mathcal{S}_L$ chosen with Algorithm 2). 
The beamforming vectors and fronthaul quantization noise powers of all the schemes are optimized by taking into account the impact of CSI errors, via the proposed robust design.
We plot the average minimum-rate $R_{\min}$ versus the relative CSI error $\rho_z$ for a cell-free MIMO system with $M=4$, $K=6$, $C_{\text{fh}} = 10$ bps/Hz, and $P_{\text{tx}}/\sigma^2 \in \{10, 20\}$ dB.
For both transmit SNR levels $P_{\text{tx}}/\sigma^2 \in \{10, 20\}$ dB, crossover points are observed between the SDMA and NOMA schemes. 
The RSMA scheme outperforms the two conventional multiple access schemes in all simulated cases even with a single $L=1$ common signal.
Since the performance gain of RSMA comes from better interference management, the gain is larger with more accurate CSI and higher SNR level.

\begin{figure}
\centering\includegraphics[width=0.7\linewidth]{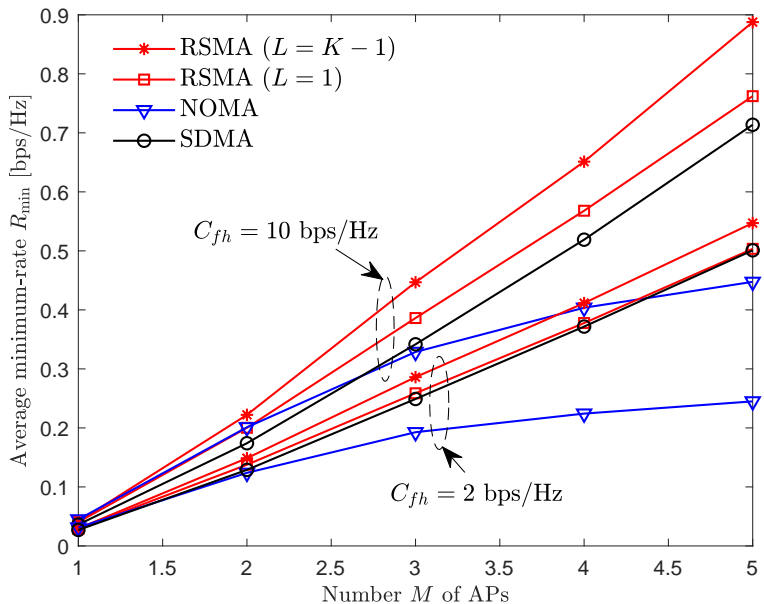}
%\vspace{-3mm}
\caption{\label{fig:vsM-various}Average minimum-rate $R_{\min}$ versus the number $M$ of APs ($K=8$, $C_{\text{fh}} \in\{2, 10\}$ bps/Hz, $P_{\text{tx}}/\sigma^2 = 15$ dB, and $\rho_z = 0.1$)}
\end{figure}

Fig. \ref{fig:vsM-various} plots the average minimum-rate $R_{\min}$ of the same schemes while increasing the number $M$ of APs for a cell-free MIMO system with $K=8$, $C_{\text{fh}} \in\{2, 10\}$ bps/Hz, $P_{\text{tx}}/\sigma^2 = 15$ dB, and $\rho_z = 0.1$.
The RSMA and SDMA schemes show similar growth rates with respect to $M$, while
the slope of NOMA curve is the smallest.
This is because there are a larger number of rate constraints that need to be satisfied for successful SIC decoding at all UEs.
Also, the performance gap among the schemes increases with the fronthaul capacity $C_{\text{fh}}$ due to the decreased quantization noise powers.

\section{Conclusion} \label{sec:conclusion}

We have studied a robust design of RSMA for the downlink of a cell-free MIMO system with finite-capacity fronthaul links and erroneous CSI.
To this end, we have considered an additive CSI error model and assumed that the CP designs the beamforming and fronthaul quantization strategies by leveraging knowledge of global nominal CSI and of the stochastic distribution of CSI errors.
To efficiently solve the problem of maximizing the minimum of expected data rates, we have used closed-form lower bound expressions and tackled the problem by an MM algorithm with rank relaxation. 
With extensive numerical results, we have observed the importance of robust design in the presence of CSI errors and how the performance gain of RSMA over conventional multiple access schemes is affected by CSI imperfection.
Among relevant future research directions, we mention the design of RSMA with binning \cite{Romero:arxiv20} and the robust design with imperfect CSI on general multicast scenarios \cite{Zhao:arxiv22}.

\end{document}